\documentclass[12pt]{article}

\usepackage{amsmath,amssymb,graphicx} %use drftcite in draftmode
\usepackage{epsf,color}
\usepackage{cite}

\newcommand{\add}[1]{{\it #1}}

\newcommand{\beq}{\begin{eqnarray}}% can be used as {equation} or  {eqnarray}
\newcommand{\eeq}{\end{eqnarray}}
\newcommand{\centeron}[2]{{\setbox0=\hbox{#1}\setbox1=\hbox{#2}\ifdim

\wd1>\wd0\kern.5\wd1\kern-.5\wd0\fi
\copy0

\kern-.5\wd0\kern-.5\wd1\copy1\ifdim\wd0>\wd1
                                       \kern.5\wd0\kern-.5\wd1\fi}}
\newcommand{\ltap}{\>\centeron{\raise.35ex\hbox{$<$}}
                               {\lower.65ex\hbox{$\sim$}}\>}
\newcommand{\gtap}{\>\centeron{\raise.35ex\hbox{$>$}}
                               {\lower.65ex\hbox{$\sim$}}\>}

\newcommand\ZZ{\hbox{\zfont Z\kern-.4emZ}}
\font\zfont = cmss10 %scaled \magstep1

\begin{document} 
\begin{titlepage}
\begin{flushright}
%{\tt hep-ph/yymmnn}
\end{flushright}

\vskip.5cm
\begin{center}
{\huge \bf
A New Custodian \\
\vskip.3cm
for a Realistic Higgsless Model}

\vskip.1cm
\end{center}
\vskip0.2cm

\begin{center}
{\bf
{Giacomo Cacciapaglia}$^{a}$,  {Csaba Cs\'aki}$^{a}$,
{Guido Marandella}$^{b}$, {\rm and}
{John Terning}$^{b}$}
\end{center}
\vskip 8pt

\begin{center}
$^{a}$ {\it Institute for High Energy Phenomenology\\
Newman Laboratory of Elementary Particle Physics\\
Cornell University, Ithaca, NY 14853, USA } \\
$^{b}$ {\it
Department of Physics, University of California, Davis, CA
95616.} \\
\vspace*{0.3cm}
{\tt  cacciapa@mail.lns.cornell.edu, csaki@mail.lns.cornell.edu, maran@physics.ucdavis.edu, terning@physics.ucdavis.edu}
\end{center}

\vglue 0.3truecm

\begin{abstract}
\vskip 3pt \noindent We present an example of a realistic Higgsless
model that makes use of alternative $SU(2)_R$ assignments for the
top and bottom quarks recently proposed by Agashe et
al. which results in an enhanced custodial symmetry. Using these new representations reduces the
deviations in the $Zb_\ell\bar{b}_\ell$ coupling to $\sim 4\%$ for a wide range of
parameters, while this remaining correction can also be eliminated
by varying the localization parameter (bulk mass) for $b_r$.

\end{abstract}

\end{titlepage}

\newpage

%\renewcommand{\thefootnote}{(\arabic{footnote})}

%%%%%%%%%%%%%%%%%%%%%%%%%%%%%%%%%%%%%%%%%%%%%%%%%%%%%%
%%%%%%%%%%%%%%%%%%%%%%%%%%%%%%%%%%%%%%%%%%%%%%%%%%%%%%
\section{Introduction}
\label{sec:intro}
\setcounter{equation}{0}
\setcounter{footnote}{0}
%%%%%%%%%%%%%%%%%%%%%%%%%%%%%%%%%%%%%%%%%%%%%%%%%%%%%%

It has been realized in the last few years that extra dimensions
allow for an alternative approach to electroweak symmetry breaking
where the Higgs decouples from the theory~\cite{CGMPT,CGPT,deconstruct}.  With
the Higgs localized on a brane one can take the limit  $\langle H
\rangle \to \infty$ while the $W$ and $Z$ masses remain finite.   In
this Higgsless limit the gauge boson masses are set by the size of
the extra dimension via Dirichlet boundary conditions (BCs) and $WW$ scattering is 
unitarized by $W$ and $Z$ Kaluza-Klein (KK) modes \cite{CGMPT,CGPT,otherunitarity,Papucci}. If the
extra dimension is warped then a bulk gauge symmetry $SU(2)_L \times
SU(2)_R \times U(1)_{X}$ acts~\cite{CGPT,ADMS} as a custodial symmetry
to ensure the correct ratio of $M_W/M_Z$ (i.e. a small $T$
parameter \cite{CCGT,Davoudiasl:2003me,Nomura,BPR}) at tree level. If the fermions are spread uniformly
through the bulk~\cite{CuringIlls,MSU1} then fermion currents are
approximately orthogonal to the gauge KK modes so
their couplings are suppressed and the $S$ parameter is
small\footnote{For another suggestion for lowering the
$S$ parameter using the condensates of a conformal field  theory (CFT)
see~\cite{HirnSanz}.}. The remaining problem in this class of
theories is how to get a large enough top quark mass without messing
up the $Z b_\ell \bar b_\ell$ coupling~\cite{CuringIlls}. This final problem
can be solved using alternative $SU(2)_R$ assignments for the top
and bottom quarks that were recently suggested by Agashe et. al.
\cite{CustodZbb}. This alternative  is suggested by a combination of custodial symmetry and a
$L \leftrightarrow R$ parity symmetry that protects the $Z b_\ell \bar b_\ell$ coupling.
This enhanced custodial symmetry suppresses  corrections to the $Zb_\ell \bar b_\ell$ vertex, but  in Higgsless models it is not sufficient by itself to reach agreement
with the experimental bounds. A cancelation, therefore, is still
necessary: we identify the simplest scenario where such a cancelation
is possible. In this scheme the left-handed (LH) top and bottom quarks are part
of a bi-doublet of $SU(2)_L \times SU(2)_R$, while the right-handed (RH)
top is a singlet and the RH bottom is part of an $SU(2)_R$
triplet. For a wide range of parameters the deviation of the
$Zb_\ell \bar{b}_\ell$ coupling is reduced to $\sim 4\%$, and this deviation can 
be eliminated by varying the bulk mass (localization parameter) of the $b_r$.

%%%%%%%%%%%%%%%%%%%%%%%%%%%%%%%%%%%%%%%%%%%%%%%%%%%%%%
%%%%%%%%%%%%%%%%%%%%%%%%%%%%%%%%%%%%%%%%%%%%%%%%%%%%%%
\section{Fermion masses in Higgsless models}
\label{sec:top}
\setcounter{equation}{0}
%\setcounter{footnote}{0}
%%%%%%%%%%%%%%%%%%%%%%%%%%%%%%%%%%%%%%%%%%%%%%%%%%%%%%

Higgsless models achieve electroweak symmetry breaking through
Dirichlet BCs for gauge fields in an extra
dimension \cite{CGMPT,CGPT,Nomura,BPR}.   The correct ratio of $W$ and $Z$
masses is predicted if the model has  a custodial symmetry~\cite{CGPT,ADMS}. In the standard model (SM) the Higgs sector has an  $SU(2)_L \times SU(2)_R$ symmetry which is broken down to a diagonal  $SU(2)_D$ custodial symmetry by the Higgs VEV. $SU(2)_L$ is a gauge symmetry in the SM,    while $SU(2)_R$ is a global symmetry that is broken by Yukawa couplings and the hypercharge gauge coupling.
(Yukawa
couplings would not break the custodial symmetry if the RH fermions were doublets of $SU(2)_R$, but then the 
$t$ and $b$ would have to be degenerate.)
In Randall-Sundrum-type models \cite{RS},  where the SM is embedded in a 5D  anti-de Sitter (AdS) space, a custodial symmetry can be achieved by  incorporating a bulk gauge symmetry  $SU(2)_L \times SU(2)_R \times
U(1)_{X}$ where $SU(2)_R \times U(1)_{X}$ is broken down to hypercharge $U(1)_Y$
by Dirichlet BCs on the UV brane. This is in accord with the AdS/CFT
correspondence, which requires that a global symmetry of the strongly coupled CFT corresponds to  a
gauge symmetry  in AdS.
On the IR brane, $SU(2)_L \times
SU(2)_R$ is broken down to an $SU(2)_D$ custodial
symmetry by Dirichlet BCs, analogously to  what happens in  the SM.

Fermion masses can be  easily generated in Higgsless models via  Dirac
masses\footnote{Note that this scenario can also be generated by a
finite Higgs VEV and a localized Higgs Yukawa coupling.}  on the IR brane
\cite{BPR,CGHST}. In 5D, fermions are vector-like so that each bulk
fermion field  contains both  LH and  RH
components: 
\beq \Psi = \left( \begin{array}{c}
\chi \\
\psi \end{array} \right)\,. \eeq 
Here, and in the remainder of the
paper, we use $\chi$ to denote the LH fermion and $\psi$ to denote
the RH fermion. We will often use subscripts $\ell$ and $r$ to denote LH and RH fermion
chirality, while L and R always indicate the gauge groups $SU(2)_L \times SU(2)_R$. A
chiral spectrum can be obtained by assigning different BCs to the two components of a 5D fermion: for
example, assigning Dirichlet BC's to the RH component $\psi$ is
enough to determine the solutions of the bulk equations of motion, and 
allows for a  LH zero mode in $\chi$. One massless
flavor thus requires two bulk fermion fields, one with a LH zero
mode and one with a RH zero mode.  The LH and RH zero modes of two fields can be removed
by adding a brane Dirac mass  for the two fields; this results in new BC's
 \cite{CGHST} which  generate  a  $z$-dependence incompatible with a zero mode.

The simplest possibility, adopted in Refs.~\cite{CGPT,CCGT,CuringIlls}, is
to embed  the LH and RH SM fermions into $SU(2)_L$ and $SU(2)_R$
bulk doublets: for instance, for the third generation quarks,
$\Psi_L=(t_L, b_L)$, $\Psi_R=(t_R, b_R)$ transforming as 
\beq
\label{oldreps}
({\bf 2},{\bf 1})_{1/6},~\,\,\,\,\, ({\bf 1},{\bf 2})_{1/6}
\eeq
 of $SU(2)_L \times
SU(2)_R \times U(1)_{X}$. For these representations, the $X$-charge can be identified as $X=(B-L)/2$. The SM zero modes can be
reproduced by the assignment of the following BC's: \beq
\begin{array}{c|cc}
 \Psi_L & \mbox{UV} & \mbox{IR} \\
 \hline
\chi_L = \vphantom{ \sqrt{\left( \begin{array}{c} \chi_{t_R} \\ \chi_{b_R} \end{array} \right) }}
 \left( \begin{array}{c} \chi_{t_L} \\ \chi_{b_L} \end{array} \right) & + & + \\
\psi_L = \left( \begin{array}{c} \psi_{t_L} \\ \psi_{b_L} \end{array} \right)& - & -
\end{array}
 \qquad
\begin{array}{c|cc}
\Psi_R & \mbox{UV} & \mbox{IR} \\
\hline
\chi_R =  \vphantom{ \sqrt{\left( \begin{array}{c} \chi_{t_R} \\ \chi_{b_R} \end{array} \right) }}
\left( \begin{array}{c} \chi_{t_R} \\ \chi_{b_R} \end{array} \right) & \left( \begin{array}{c} - \\ \mp \end{array}\right) & -  \\
\psi_R = \left( \begin{array}{c} \psi_{t_R} \\ \psi_{b_R} \end{array} \right)  & \left( \begin{array}{c} + \\ \pm \end{array}\right) & +
\end{array}
\eeq where $+$ stands for a Neumann BC and $-$ stands for a
Dirichlet BC, and  (in the absence of boundary mass terms) the LH
and RH components of a field must always have opposite BCs. Note
that $SU(2)_R$ is broken on the UV brane, so that we can assign
different BC's to fields in the $SU(2)_R$ doublet: this is the
origin of the two possible choices for $\Psi_R$. The difference in
the two cases is the presence, or not, of a zero mode for the
$b_r$: in the latter case one can add another $SU(2)_R$ doublet $\Psi'_R$
flipping the BC's between top and bottom quarks so as to get a RH
$b$ zero mode.

A localized mass term on the IR brane of the form: 
\beq
\label{eq:fermionmass} M \left( \chi_L \psi_R + \chi_R \psi_L
\right) + h.c. \eeq 
replaces \cite{CGHST} the Dirichlet BC's on the
IR brane ($\psi_L=0$  and $\chi_R=0$) with: \beq \label{eq:massBC}
\psi_L = M R' \psi_R\,, \quad \chi_R = - M R' \chi_L\,. \eeq It is clear
from these BC's that it is not possible to obtain an arbitrarily
large top quark mass by increasing $M$. The reason is that in the large
$M$ limit the BC's are equivalent to: \beq \label{eq:BClimit} \chi_L
= 0\,, \quad \psi_R = 0\,, \eeq which just amounts to flipping the
BC's on the IR brane, so that the top  quark corresponds to the first KK
mode of a field with $(+,-)$ BC's whose mass is set by the radius of
the extra dimension (rather than the localized Dirac mass).

Note that in order to preserve custodial symmetry, the mass term
(\ref{eq:fermionmass}) couples to both   top and bottom quarks. If
the RH $b$ lives in $\Psi_R$, the bottom mass can be suppressed with
respect to the top mass by a large kinetic term on the UV brane for
$\psi_{b_R}$. In the case of different UV BCs for $\psi_{t_R}$ and
$\psi_{b_R}$, the mass term (\ref{eq:fermionmass})  will not generate the bottom mass, due to
the absence of a zero mode for $\psi_{b_R}$.

Despite the seeming ease of introducing fermion masses, they in fact
pose the main challenge  to Higgsless models. First of all, the
necessity of having the light fermions spread in the bulk, due to
electroweak precision corrections, is potentially dangerous from the
point of view of flavor physics~\cite{CuringIlls} without some
additional flavor symmetry. The second, more difficult  challenge
comes from  the third generation of quarks: there is a tension
between having a heavy top quark and small corrections to the
couplings of the LH $b$ with the $Z$ boson. Schematically the large
corrections to $Z b_\ell \bar b_\ell$ have two origins. First of all, in order
to enhance the $t$ mass, one has to localize the $t$ as close as
possible to the IR brane. However, electroweak symmetry breaking is
also localized there which distorts the wave functions of the
$W$ and $Z$, thus modifying the coupling of any field
localized nearby. The second source of deviations is the presence of
a massive LH $b^\prime$ quark (electric charge $-1/3$) in the
$SU(2)_R$ doublet that contains the RH $t$: this fermion
mixes with the  $b$ via the IR Dirac mass responsible for
the $t$ mass, however it has hypercharge $-1/3$ instead of the usual
$+1/6$. This mixing is a direct consequence of the custodial
symmetry. Unfortunately, if the $t$ is not localized extremely close to the IR
brane, the heaviness of the $t$ requires the Dirac mass term (\ref{eq:fermionmass})  on the
IR brane to be very large, so that a large mixing is generated.
These two sources together generate large deviations for  the entire
parameter space.

One possible solution to this problem would be to increase the mass
scale on the IR brane, without conflicting with the unitarity bound
in the gauge sector. In Ref.~\cite{twobranes} this idea was
realized in a model with  two Randall-Sundrum throats where light fermions and
the third generation separately couple to  symmetry breaking on two different
IR branes. However, in order for the gauge sector to be not
sensitive to the large scale of the top-sector IR-brane, the top is
necessarily strongly coupled to a Higgs and/or resonances living
on the top-brane. Although the gauge sector is screened by a weak
loop from those strong coupling effects, the top sector is not
calculable.\footnote{An analogous separation of scales for fermion
and gauge boson masses in the deconstructed version of the model has
been proposed in~\cite{MSU}.}

Here we will pursue another possibility that allows for a perturbative top
sector: an alternative realization of custodial
symmetry \cite{CustodZbb} that does not require the presence of a
$b$-like particle in the bulk field containing the RH
$t$. Furthermore, $b_\ell$ couples to the diagonal combination of
$SU(2)_L \times SU(2)_R$, that is flat near the IR brane. In  this way
both of the sources for the deviation of the $Zb_\ell \bar b_\ell$ coupling
are suppressed.

%%%%%%%%%%%%%%%%%%%%%%%%%%%%%%%%%%%%%%%%%%%%%%%%%%%%%%%%%%%%%%%%%%
\section{An alternative  realization of custodial symmetry} \label{sec:alternative}
%%%%%%%%%%%%%%%%%%%%%%%%%%%%%%%%%%%%%%%%%%%%%%%%%%%%%%%%%%%%%%%%%%

In \cite{CustodZbb}, the authors identified representations of $SU(2)_L \times SU(2)_R$
 that have an enhanced custodial symmetry which can protect the coupling of the $Z$ to a
given fermion $\psi$ from non-universal corrections. Let us briefly
summarize the argument of  \cite{CustodZbb}. One assumes a
beyond the SM  sector with an 
$O(4) \sim SU(2)_L \times SU(2)_R \times P_{LR}$ symmetry, where $P_{LR}$
is the discrete parity interchanging the two $SU(2)$'s.  $O(4)$ is
then broken down to $O(3) \sim SU(2)_D \times
P_{LR}$.  If $\psi$ is a $+1$ eigenstate of $P_{LR}$ then
\beq \label{eq:sym1}
T_L=T_R, \;\;\;\;\;\; T_R^3=T_L^3
\eeq
and since the $Q_{L+R}$ charge is protected by $SU(2)_D$ (so it is not renormalized) and shifts in $Q_L$ and $Q_R$ must be equal we have
\beq
\delta Q_L+\delta Q_R=0, \;\;\;\;\;\; \delta Q_L=\delta Q_R,
\eeq
so the charges are individually protected ($\delta Q_L=0$), and the $W^3_L$ coupling is unchanged.
In addition fermions with 
\beq \label{eq:sym2}
T_L^3 = T_R^3 =0
\eeq
 cannot couple to $W^3_L$ at all.

The field we are interested in is the most precisely measured of the
third generation,  i.e. the LH $b$ quark. We need the field containing
$b_\ell$  to satisfy  (\ref{eq:sym1}). The minimal choice is to embed $b_\ell$  in a
bi-doublet of $SU(2)_L\times SU(2)_R$. Consequently, the RH
fermions can be either singlets or triplets of $SU(2)_L$ and/or
$SU(2)_R$. Some possible choices (labeled by $SU(2)_L
\times SU(2)_R \times U(1)_{X}$ quantum numbers) are:
\add{\beq
\begin{array}{ll}
({\bf 1},{\bf 1})_{2/3} \supset t_{r}  \;\;\;\; & ({\bf 1},{\bf 3})_{2/3}\; \mbox{{\rm  and/or  }}\; ({\bf 3},{\bf 1})_{2/3} \supset t_{r}\,, b_{r}\\
({\bf 1},{\bf 1})_{-1/3} \supset b_{r} & ({\bf 1},{\bf 3})_{-1/3}\; \mbox{{\rm  and/or  }}\; ({\bf 3},{\bf 1})_{-1/3}\supset t_{r}\,, b_{r}~.
\end{array} \eeq}

For instance, the third generation of quarks could be obtained from
the following representations:
\beq
\label{newreps}
\begin{array}{cll}
\Psi_L &\sim
({\bf 2},{\bf  2})_{2/3} &\supset (t_\ell ,b_\ell )\\
t_R & \sim ({\bf 1},{\bf 1})_{2/3} & \supset t_{r}\\
\Psi_R & \sim({\bf 1},{\bf 3})_{2/3} & \supset b_{r}
\end{array} \eeq

This is one of the two cases where the realization of the custodial
symmetry identified in~\cite{CustodZbb} protects the $Z b_\ell\bar b_\ell$
couplings\footnote{In the other case, $t_{r}$ is embedded in a
$({\bf 1},{\bf 3})_{2/3} \oplus ({\bf 3},{\bf 1})_{2/3}$. We focus on the
former scenario because it involves fewer bulk fields: however, all
of the following discussion can be applied to the latter case as
well.}. Note that the field containing $t_{r}$ does not contain
any field that can mix with $b_\ell$: in other words, the $t$ mass
will not induce any mixing in the $b$ sector. The $b$ mass is
generated by a different (and smaller) Dirac mass.

Notice that in order for the enhanced custodial symmetry to protect the $b$, $\Psi_R$ should be completed to a full $O(4)$ representation:  $\Psi_R = ({\bf 3},{\bf 1})_{2/3} \oplus ({\bf 1},{\bf 3})_{2/3}$.
This is because in the $SU(2)_R$ triplet $({\bf 1},{\bf 3})_{2/3}$, the component with $T_R^3 = -1$ has the quantum numbers of a $b$.
We can choose the BC's for this field such that a zero mode is only present in the RH component, while the LH modes are all massive.
However, the Dirac mass needed to get the $b$ mass mixes it with the LH component living in $\Psi_L$. 
Thus, the LH $b$  lives partly in $({\bf 1},{\bf 3})_{2/3}$.
In order to protect the $Z b_\ell \bar b_\ell$ vertex, we need to complete the representation with a $SU(2)_L$ triplet: its component with $T_{L}^3 = -1$ will also mix with $b_L$, and it will cancel out the contribution of the $SU(2)_R$ triplet (component with $T_R^3=-1$) to $Z b_\ell \bar b_\ell$.
Usually, since the $b$ quark mass is small, it can be neglected and completing the RH fields to full $O(4)$ representations is irrelevant. However, in the context of the Higgsless model, due to possible different localizations of $\Psi_L$ and $\Psi_R$ the Dirac mass needed to get the $b$ mass can be quite large. Thus it might seem that completing  $\Psi_R$ to a full $O(4)$ representations could be crucial. We will come back to this point in the next section.

For the time being let us consider $\Psi_R$ to be a $SU(2)_R$
triplet and write, in 5D components, 
\beq \Psi_L = \left(
\begin{array}{cc} q_L & Q_L
\end{array} \right) = \left( \begin{array}{cc}
t_L & X_L \\
b_L & T_L
\end{array} \right), \qquad 
 \Psi_R = \left( \begin{array}{c}
X_R\\ T_R \\ b_R
\end{array} \right), \qquad
t_R~, \eeq where all these fermion fields are bulk fields. For example,
$t_R$  contains both LH  and RH components $\chi_{t_R}$ and
$\psi_{t_R}$. Notice also that the charges of the extra fields are
$Q[T] = 2/3$ and $Q[X] = 5/3$, so that $T$  will mix with $t$.
The quantum numbers, and BC's, of the various fields are shown in Tables~1 and 2.
%\ref{tab:quantumnumbs}.
As usual, $Y=T^3_{R}+X$,  $Q=T^3_{L}+Y$, and the BC's
ensure that the only zero modes correspond to   SM fields.

\begin{table}[htb]
 \label{tab:quantumnumbs}
\begin{center}
\begin{tabular}{c|rr}
& $T^3_{L}$ & $T^3_{R}$ \\ \hline
$X_L$ & $1/2$ & $1/2 $ \\
$T_L$ & $-1/2$ & $1/2$ \\
$t_L$ & $1/2$ & $-1/2$ \\
$b_L$ & $-1/2$ & $-1/2$ \\
$X_R$ & $0$ & $1$ \\
$T_R$ & $0$ & $0$ \\
$b_R$ & $0$ & $-1$ \\
$t_R$ & $0$ & 0
\end{tabular} \hspace{2 cm}
\begin{tabular}{c|cc}
 & UV & IR\\
\hline
$\chi_{q_L}$ & $+$ & $+$ \\
$\chi_{Q_L}$  & $-$ & $+$\\
$\psi_{X_R}$& $-$&$+$ \\
$\psi_{T_R}$&$-$&$+$\\
$\psi_{b_R}$&$+$&$+$\\
$\psi_{t_R}$&$+$&$+$
\end{tabular}
\end{center}
  \caption{{Quantum numbers of the bulk fields,
and Table 2: 
 BC's of the bulk fields.}}
\end{table}
\setcounter{table}{2}

The mass terms can be written as: 
\beq 
\mathcal{L}_{IR} =
M_3\left[\frac{1} {\sqrt{2}} T_R\,  \left(t_L + T_L \right) +  b_R b_L +   X_R X_L\right]
+ \frac{M_1}{\sqrt{2}} \; t_R\, \left(t_L -T_L \right) +h.c. 
\eeq 
Under the unbroken $SU(2)_D$ on the IR brane,
$(t_L -T_L)/\sqrt{2}$ is the singlet component of $\Psi_L$, while
$(X_L,(t_L +T_L)/\sqrt{2},b_L)$ is the triplet. Note that the $t_L$ is
mixed with the massive $T_L$, so that a larger Dirac mass is
required than was the case for the representations (\ref{oldreps}). This implies that the $t$ has to be localized closer to
the IR brane than in the previous case. Moreover, the $b$ only
feels the smaller  $M_3$ term, so the $b$ wave function is not distorted
as much as the $t$.

%%%%%%%%%%%%%%%%%%%%%%%%%%%%%%%%%%%%%%%%%%%%%%%%%%%%%%%
\section{Results for  Higgsless model}
%%%%%%%%%%%%%%%%%%%%%%%%%%%%%%%%%%%%%%%%%%%%%%%%%%%%%%%

\begin{figure}
  \centering
  \includegraphics[width=0.48\textwidth]{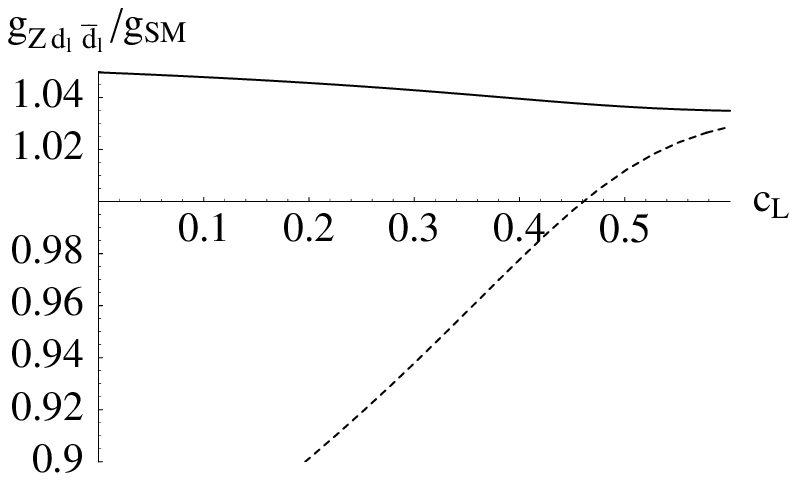}
  \includegraphics[width=0.48\textwidth]{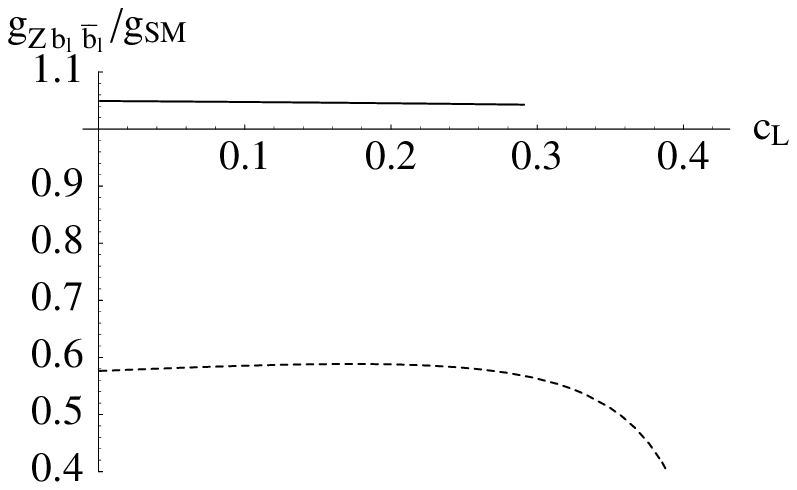}
  \caption{Deviations in the coupling  of the down-type quarks as a function of the bulk mass $c_L$ using the alternative realization of custodial symmetry, where  $\Psi_L=({\bf{2,2}}), \Psi_R = ({\bf{1,3}}), t_R=({\bf{1,1}})$ (continuous line) and the original case, where $\Psi_L=({\bf{2,1}}), \Psi_R = ({\bf{1,2}})$ (dashed line). In the left panel, we consider a light quark (first and second generation). In the right panel, we show the $b$ quark, with $c_R^t = 0$, while neglecting the $b$ mass. }
  \label{fig:nombottom}
\end{figure}

We first consider applying the new representation (\ref{newreps}) to the light fermions,
whose masses are so small  that mass dependent corrections can be neglected. In
Fig.~\ref{fig:nombottom} we show the coupling of the LH down-type
quark $d_\ell$ with the $Z$ with respect to the SM coupling: for the alternative
representation (\ref{newreps}), the $Z d_\ell \bar d_\ell$ coupling deviates by  $+4$\% to $+5$\%,
independently of the bulk mass of the $\Psi_L$ fields. This correction
has a universal origin which is equivalent  to a large $S$ parameter \cite{CuringIlls,diguise}. For
 light fermions in the representations (\ref{oldreps}), this can be
compensated by a suitable value of the bulk mass\footnote{In units of the AdS scale $1/R$.} $c_L \simeq 0.45$, as
in~\cite{CuringIlls}. With the alternative
representation it is not possible to cancel the correction,
since the coupling is now protected by the enhanced symmetry and
does not depend on the localization of the fermions. This fact
 implies that the alternative representation (\ref{newreps}) cannot be used
for the light fermions in Higgsless models  due to this large
vertex correction.
There is a very clear way of understanding this in
terms of KK modes with electroweak symmetry breaking treated as a
perturbation: the correction is generated by a tree level graph
where the $b_\ell$ emits a KK mode of the $Z$, that mixes with the zero
mode due to electroweak symmetry breaking BC's on the IR brane. 
There are three different kind of gauge KK modes: those from $SU(2)_L$, those from  the combination of $SU(2)_R$ and $U(1)_X$ that is unbroken on the UV brane (which couples through hypercharge $Y$), and those from the broken combination.
In the limit of flat  fermions ($c_L=0.5$), only the latter KK modes will contribute and give the few \% deviation we observe, because they do not have a zero mode and are thus not orthogonal to the flat wave function.
When  $c_L\ne 0.5$, the other modes will also contribute, but, due to the enhanced symmetry, they will cancel so as to keep the deviation constant.
On the other hand, in the original case (\ref{oldreps}), we can tune the parameters such that  the contributions completely cancel each other.

Next, we neglect the $b$ mass ($M_3 = 0$), and consider only the
$t$ mass. As already mentioned, despite the fact that $\Psi_R$ should be completed to a full $O(4)$ representation, if only $M_1$ is considered $Z b_\ell \bar b_\ell$ remains protected: the $t$ mass does
not affect the coupling of the $b$, and the correction remains at $+4$
to $+5$\%. In the old case  (\ref{oldreps}), the corrections were typically $\sim -40$\%, due to
the large mixing of the $\chi_{b_L}$ with a LH field contained in the $\Psi_R$
doublet: thus, the new symmetry allows for a significant reduction in the
correction, but it is still not enough to be compatible with the
experimental bounds. Another difference between
the two cases is that with the new representation (\ref{newreps}) the approximate zero modes in 
$\Psi_L$ have
to be more localized toward the IR brane in order to produce the observed   $t$
mass: in fact, $c_L  \lesssim 0.3$ is required, compared with $c_L \lesssim 0.45$
in the old case. As we mentioned before the reason behind this is
the  mixing of the $t$ and $T$ fields. The
results are shown in Fig.~\ref{fig:nombottom}.

\begin{figure}[t]
\begin{center}
\includegraphics[width=0.5\textwidth]{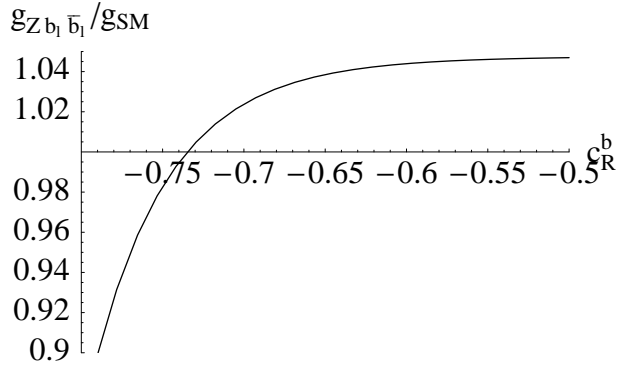}
\caption{Deviations in the coupling $Z f_\ell \bar f_\ell$ as a function of the  bulk mass $c^b_R$ of $SU(2)_R$ triplet containing the $b_R$ using the alternative  realization of custodial symmetry. The other bulk masses have been set to  $c_L=0.1, \; c^t_R=0$.}
\label{fig:mbottom}
\end{center}
\end{figure}

Next we  analyze the same corrections, taking into account the
$b$ mass. As we said in Sec.~\ref{sec:alternative} if $M_3$ is considered, and $\Psi_R$ is an $SU(2)_R$ triplet, neither (\ref{eq:sym1}) nor (\ref{eq:sym2}) are satisfied. So, we expect deviations in the $Z b_\ell \bar b_\ell$ coupling. Notice that if we completed $\Psi_R =  ({\bf 3},{\bf 1})_{2/3} \oplus ({\bf 1},{\bf 3})_{2/3} $ nothing would change with respect to the  case with $M_3=0$: the deviation would remain between 4 and 5\% since the enhanced custodial symmetry would be at work. If a
sizeable Dirac mass is necessary to fit the $b$ mass, then the
LH mode in $b_R$ (the triplet component with electric
charge $-1/3$ but with hypercharge 2/3) mixes with the LH mode in $b_L$ and
can cancel the correction to $Z b_\ell \bar b_\ell$.  The
size of the Dirac mass needed to get the $b$ mass crucially
depends on the bulk mass parameter $c^b_R$ for the field $\Psi_R$.
If this is large and negative, i.e. the approximate zero mode in $\Psi_R$ is localized near the
UV brane, the overlap between $b_L$ and $b_R$ is small, and the
Dirac mass is of order the IR scale $1/R'$. Such a big Dirac mass
induces a sizeable mixing between the LH modes living in $b_L$ and
$b_R$, giving a negative correction to the $Z b_\ell \bar b_\ell$
coupling. Note that the sign of this correction goes in the right
direction to agree with the SM.
In Fig. \ref{fig:mbottom} we plot the
correction to the $Z b_\ell \bar b_\ell$ coupling as a function of $c^{b}_R$.
In this plot, we choose $c_L = 0.1$ while the bulk mass parameter
corresponding to $t_R$ has been set to $c^t_R=0$. In the notation of ref. \cite{CGPT}, we have set the AdS scale to $1/R=10^{8}$ GeV, while the IR scale is determined by $M_Z$, $G_F$ and $\alpha(M_Z)$ to be $1/R' = 282$ GeV. This plot has a
mild dependence on $c_L$ and $c^t_R$. For $c^b_R \sim -0.75$, the
two contributions cancel out, and the coupling agrees with the SM to
the required accuracy. To summarize, in the scenario where the approximate zero modes of $t_L,
t_R, b_L$ are localized near the IR brane ($c_L \lesssim 0.3, c^t_R
\simeq 0$), and $b_R$ is localized close to the UV brane ($c^{b}_R
\simeq -0.75$), allows for a fully realistic Higgsless model: the $t$
mass can be obtained, while corrections to the $Z b_\ell \bar b_\ell$
coupling can be made arbitrarily small.

We can now examine the
couplings  of the third generation  quarks and  gauge bosons.
Setting $c_L=0.1, \; c^t_R=0, \; c^b_R = -0.73$, we find the values shown in Table 3.
%\ref{tab:couplings}.

\begin{table}[hb]
 \label{tab:couplings}
\begin{center}
% \begin{tabular}{c|cc}
%  & Higgsless & frac. of SM \\  \hline \hline
% $\vphantom{\sqrt{F^{F^F}}} Z b_\ell \bar b_\ell$ & $-0.316$ & 1.004  \\
% $Z b_r \bar b_r $ & 0.057 &  0.993 \\
% $ Z t_\ell \bar t_\ell$ & 0.1186   & 0.461 \\
% $ Z t_r \bar t_r$ &$-0.21$ & 1.908 \\
% $W t_\ell \bar b_\ell $ & 0.2849 & 0.874 \\
% $ W t_r \bar b_r$  &$3 \cdot 10^{-4}$ &-
% \end{tabular}
\begin{tabular}{c|l}
 & frac. of SM \\  \hline \hline
$\vphantom{\sqrt{F^{F^F}}} Z b_\ell \bar b_\ell$ & 1.004  \\
$Z b_r \bar b_r $ &  0.993 \\
$ Z t_\ell \bar t_\ell$  & 0.461 \\
$ Z t_r \bar t_r$ & 1.908 \\
$W t_\ell \bar b_\ell $ & 0.862 \\
$ W t_r \bar b_r$  &$3 \cdot 10^{-4}  \; g_{W t_\ell \bar b_\ell}$
\end{tabular}
\end{center}
  \caption{{The couplings of the $t$ and $b$ to $W$ and $Z$ in a Higgsless model. We have taken the AdS scale to be $1/R=10^8$ GeV, the compactification scale to be $1/R'=282$ GeV.}}
\end{table}

Notice that the deviations in the couplings of the $Z$ to the $t$
lead to a modification of the SM one-loop contribution to the $Z b_\ell
\bar b_\ell$ coupling.  Analogously, choosing for $\Psi_R$ to be an incomplete $O(4)$ representation will produce, through loop effects, a breaking of the $L\leftrightarrow R$ symmetry. This might feed back into corrections to the $Z b_l \bar b_l$ coupling. All these corrections are at most around 1 \% and a
suitable value of $c^b_R$ can always compensate it, as it is clear
from Fig. \ref{fig:mbottom}. On the other hand the deviation of the
coupling with the $W$ is small enough not to affect  present
measurements. The $W t_\ell b_\ell$ coupling will be measured by
single $t$ production, first at the Tevatron and eventually at the
 LHC with a precision around 5\%  \cite{Alwall:2006bx}.

We now summarize the features of a realistic Higgsless model at a fairly typical point in the parameter space. In Table \ref{tab:summary} we give the spectrum of the first KK excitations of top and bottom quarks, the gauge bosons, and their couplings to the SM fermions as a fraction of the analogous SM coupling. We have set the AdS scale to be $1/R=10^8$ GeV, the IR scale to be $1/R'=282$ GeV, and the ratio between the $U(1)$ and the $SU(2)$ 5D gauge couplings to be $g_5 =0.66 \; (R \log R'/ R)^{1/2}, \tilde g_5 = 0.42 \; (R \log R'/ R)^{1/2} $. The bulk mass for the light fermions is $c_L=0.46$, which makes  the $S$ parameter vanish. The parameters for the third generation are as described above: $c_L=0.1, \; c^t_R=0, \; c^b_R = -0.73$.

\begin{table} \label{tab:summary}
  \centering
  \begin{tabular}{c|l}
    $M_{t'}$ & 450 GeV \\
    $M_{b'}$ & 664 GeV \\
    $M_{W'}$ & 695 GeV  \\
    $M_{Z'}$ & 690 GeV  \\
    $M_{Z''}$ & 714 GeV  \\
    $M_{G'}$ & 714 GeV  \\
    $g_{W' u \bar d}$ & $0.07 \; g $  \\
    $g_{Z' q \bar q}$ & $0.14 \; g_{Z q \bar q}$  \\
    $g_{G' q \bar q}$ & $0.22 \; g_c$ 
   \end{tabular} \hspace{1cm}
  \begin{tabular}{c|l}
    $g_{Z' t_L \bar t_L}$ & $1.83 \; g_{Z t_L \bar t_L} $   \\
    $g_{Z' t_R \bar t_R}$ & $4.02 \; g_{Z t_R \bar t_R}$ \\
   $g_{Z' b_L \bar b_L}$ & $3.77 \; g_{Z b_L \bar b_L}$  \\
    $g_{Z' b_R \bar b_R}$ & $0.26 \; g_{Z b_R \bar b_R}$  \\
    $g_{ZWW}$  & $1.018 \; g\, c_W$\\
    $g_{ZZWW}$  & $1.044 \; g^2 c_W^2$\\
    $g_{WWWW}$ &  $1.032 \; g^2$ \\
    $g_{Z'WW}$  & $0.059 \; g\, c_W$\\
    $g_{ZW'W}$  & $0.051 \; g\, c_W$
  \end{tabular}
  \caption{Summary of the realistic Higgless model for the following choice of parameters: $1/R=10^8$ GeV, $1/R'=282$ GeV, $g_5 =0.66 \; (R \log R'/ R)^{1/2}, \tilde g_5 = 0.42 \; (R \log R'/ R)^{1/2}$. For the light fermions $c_L=0.46$, while for the third generation $c_L=0.1,\; c_R^t=0, \; c_R^b=-0.73$.}
\end{table}

For the chosen value of $R$ the gauge KK modes are close to the Tevatron bounds, and so could conceivably be discovered  or ruled out soon.  As the Tevatron bounds   move up, they effectively raise the bound on $1/R$, and all the gauge KK masses move up \cite{CuringIlls}.
If the $W^\prime$ and $Z^\prime$ KK masses go above 1 TeV, then $WW$ scattering becomes strongly coupled and the Higgsless scenario is no longer calculable \cite{Papucci,Davoudiasl}.

It is also interesting to notice the deviations in the 3- and 4-gauge
boson couplings:  those deviations are one of the main signatures of
 Higgsless models and do not depend on the details of the fermion sector. A
plot of such deviations as a function of the parameters of the model
in the gauge sector can be found in~\cite{LesHouches}.  Triple gauge boson couplings could
be measured at one part in a thousand at a future ILC.

%%%%%%%%%%%%%%%%%%%%%%%%%%%%%%%%%%%%%%%%%%%%%%%%%%%%%%%
\section{Conclusions}
%%%%%%%%%%%%%%%%%%%%%%%%%%%%%%%%%%%%%%%%%%%%%%%%%%%%%%%
We have found an example of a realistic Higgsless
model that makes use of alternative $SU(2)_R$ assignments for the
top and bottom quarks recently proposed by Agashe et
al. \cite{CustodZbb}. Using these $L\leftrightarrow R$ symmetric  representations reduces the
deviations in the  $Zb_\ell\bar{b}_\ell$ coupling, even when the  large   $t$ quark mass is taken
into account.  If the bottom quark mass is also taken into account, then for a particular value of  the $b_R$ bulk mass  
the $Zb_\ell\bar{b}_\ell$ coupling is in agreement with  precision electroweak measurements.
   
   One can take the model described here as an existence proof for the possibility of Higgsless extra-dimensional models actually being realized in our Universe. They also can provide a reasonable benchmark for comparison with experiment. It would still be interesting to find more elegant Higgsless models where
  the agreement with precision electroweak measurements is guaranteed by additional symmetries rather than parameter adjustment. Nature, however is the ultimate arbiter of what is sufficiently elegant,
 and hopefully in a few years the LHC will reveal the scenario that Nature has chosen.

%%%%%%%%%%%%%%%%%%%%%%%%%%
\section*{Acknowledgements}
We thank   Kaustubh~Agashe, Roberto Contino, Alex Pomarol, and Marco Serone for useful
discussions and comments. C.C. and J.T.  thank the Galileo Galilei
Institute for Theoretical Physics for their hospitality  and the
INFN for partial support during the completion of this work. The
research of G.C. and C.C. is supported in part by the DOE OJI grant
DE-FG02-01ER41206 and in part by the NSF grants PHY-0139738  and
PHY-0098631. G.M. and J.T. are supported by the US Department of
Energy grant DE-FG02-91ER40674.

\end{document}